\documentclass[12pt]{article}
\newcommand{\x}{arXiv:}

\newtheorem{proposition}{Proposition}

\usepackage[nosort]{cite}
\usepackage{graphicx}
\usepackage[font={scriptsize,singlespacing}]{caption}
\usepackage{epstopdf}
\usepackage{amsmath}
\usepackage{amssymb}
\usepackage{subfigure}
\usepackage{hyperref}
\usepackage{url}
\usepackage{xcolor}
\usepackage{color}

\def\sideremark#1{\ifvmode\leavevmode\fi\vadjust{\vbox to0pt{\vss
 \hbox to 0pt{\hskip\hsize\hskip1em
 \vbox{\hsize2cm\tiny\raggedright\pretolerance10000
 \noindent #1\hfill}\hss}\vbox to8pt{\vfil}\vss}}}%
\definecolor{amaranth}{rgb}{0.9, 0.17, 0.31}
\definecolor{purple(munsell)}{rgb}{0.62, 0.0, 0.77}
\definecolor{americanrose}{rgb}{1.0, 0.01, 0.24}
\definecolor{palatinateblue}{rgb}{0.15, 0.23, 0.89}
\definecolor{royalblue(web)}{rgb}{0.25, 0.41, 0.88}
\definecolor{hanpurple}{rgb}{0.32, 0.09, 0.98}
\definecolor{beaublue}{rgb}{0.74, 0.83, 0.9}
\definecolor{carminered}{rgb}{1.0, 0.0, 0.22}
\definecolor{brightpink}{rgb}{1.0, 0.0, 0.5}
\hypersetup{ linktoc=all,
    colorlinks, linkcolor={palatinateblue},
    citecolor={brightpink}, urlcolor={purple(munsell)}
}

\begin{document}
\thispagestyle{empty}
\begin{center}

\null \vskip-1truecm \vskip2truecm

{\Large{\bf \textsf{Charge Loss (or the Lack Thereof) for AdS Black Holes}}}

\vskip1truecm

{\large \textsf{
}}

\vskip0.1truecm
\textbf{\textsf{Yen Chin Ong}}\\
{\small \textsf{Leung Center for Cosmology and Particle Astrophysics \&\\ Graduate Institute of Astrophysics,  \\National Taiwan University,
Taipei 10617, Taiwan}\\
\textsf{Email: ongyenchin@member.ams.org}}\\

\vskip0.4truecm
\textbf{\textsf{Pisin Chen}}
{\small \textsf{\\Leung Center for Cosmology and Particle Astrophysics \& \\ Graduate Institute of Astrophysics \& Department of Physics,\\  National Taiwan University,
Taipei 10617, Taiwan}\\
\textsf{Email: pisinchen@phys.ntu.edu.tw}}\\

\end{center}
\vskip1truecm \centerline{\textsf{ABSTRACT}} \baselineskip=15pt

\medskip
The evolution of evaporating charged black holes is complicated to model in general, but is nevertheless important since the hints to the Information Loss Paradox and its recent firewall incarnation may lie in understanding more generic geometries than that of Schwarzschild spacetime.  
Fortunately, for sufficiently large asymptotically flat Reissner-Nordstr\"om black holes, the evaporation process can be modeled via a system of coupled linear ordinary differential equations, with charge loss rate governed by Schwinger pair-production process. The same model can be generalized to study the evaporation of AdS Reissner-Nordstr\"om black holes with flat horizon. It was recently found that such black holes always evolve towards extremality since charge loss is inefficient. This property is completely opposite to the asymptotically flat case in which the black hole eventually loses its charges and tends towards Schwarzschild limit. We clarify the underlying reason for this different behavior.

\newpage
\addtocounter{section}{1}
\section* {\large{\textsf{1. Charged Black Holes Amidst Paradoxes}}}

The discovery that black holes radiate \cite{Hawking1, Hawking2} raised the important question about the fate of the information that falls through the horizon. A pure state that collapses to form a black hole ends up as thermal radiation of mixed state after the black hole evaporates away [see however, \cite{myers}]. Such violation of unitarity is argued to be a dire consequence from the point of view of quantum theory [although not everyone agrees \cite{kn:Wald}]. 

Many ideas had since been proposed to retrieve information from black holes. One rather commonly accepted idea is that the information is not truly lost, but highly scrambled \cite{SekinoSusskind} and entangled among the Hawking radiation. 
The late time Hawking radiation subsequently purifies the earlier Hawking radiation, thus recovering the initial pure state \cite{page0, suss1}. 
Indeed, it has been proposed that by collecting and then running the collected radiation through a powerful quantum computer, the scrambled information can be decoded \cite{haydenpreskill} [see however \cite{suss1b,suss3,suss4}, and discussion below]. Despite this, the Information Loss Paradox was not resolved. On the contrary, the issue is now plagued with even more confusions when it was raised that allowing information to be recovered in such a manner [apparently] causes an equally unpalatable consequence --- spacetime at the horizon ceases to be vacuum and is in fact shrouded by ``firewall'' \cite{amps,kn:apologia} [see also \cite{sam}]. This is in conflict with what we expect from general relativity --- there should be ``no drama'' at the horizon, especially if the black hole is so large that the curvature there is negligibly small. 

In the literature, most of the discussions about information loss and firewall are centered around [asymptotically flat] neutral black holes, mainly because these are the simplest. However, such black holes inevitably get extremely hot near the end of their evaporation. At such temperature, physics is poorly understood as effective field theory begins to fail. Furthermore, since black holes can and do pick up electrical charges and/or angular momentum even if it starts with perfectly neutral and static initial configuration [charged particles are created in the vicinity of the hole by quantum fluctuation --- in fact, by the Hawking process], it is important to study more generic black hole geometries than pure Schwarzschild manifold. 

In a recent work \cite{omc}, it is argued that the best arena to study evaporating charged black holes is anti-de Sitter [AdS] space, because it is also in the context of AdS/CFT correspondence \cite{kn:Mal} that maintenance of unitarity of black hole evaporation is most evident. In particular, the idea is to study charged black holes with \emph{flat} horizon, that is, the horizon has either planar or toral topology. [In 4-dimensional spacetime, a planar horizon has trivial $\mathbb{R}^2$ topology, while toral horizon has quotient topology $\mathbb{T}^2=\mathbb{R}^2/\mathbb{Z}^2$.] These \emph{charged flat black holes} have the virtue that they are dual to field theory that behaves a lot like Quark-Gluon Plasma [QGP] \cite{kn:AdSRN}, and as such constitutes one of the most well-understood system of quantum gravity. We find that these charged black holes -- in the regime of validity of our model -- always evolve towards extremal limit. This allows us to deduce that such black holes are always destroyed in one way or another as it becomes sufficiently cold, and thus Harlow-Hayden proposal \cite{kn:HH, suss2} [that there is not enough time to decode Hawking radiation before the black hole disappears] can be made to work. [Whether this resolves the Firewall issue is of course another question].   

\begin{figure}[!h]
\centering
\includegraphics[width=0.70\textwidth]{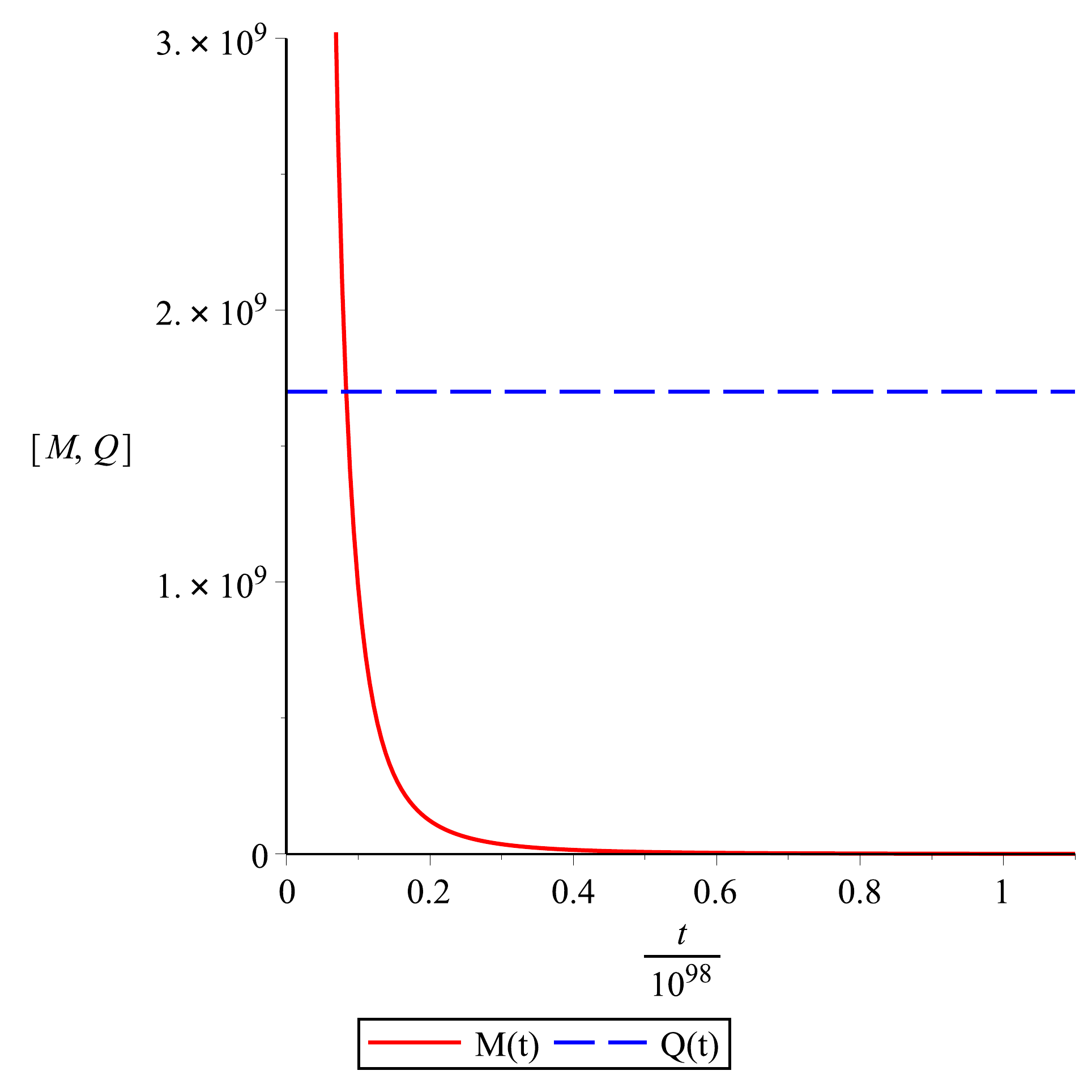}
\caption{The evolution of mass and charge of a generic toral black hole with large $L$. Time $t$ is measured in years. Note that we are allowed to have $Q>M$ since the extremal black hole does \emph{not} satisfy $Q=M$ [see section 3]. The charge $Q$ is not strictly constant, but drops by an amount too small to be noticeable at this scale. \label{1}} \end{figure}

The reason that charged flat black holes always evolve towards extremality is simple -- charge loss is inefficient. This can be seen by numerically plotting the evolution of charge as the function of time. For very generic initial conditions within the regime of validity of the model [AdS length scale $L \gg 10^8$ cm], we can see that while mass loss is quite evident, electric charge seems to be held constant throughout the evaporation history [see Fig.(\ref{1})]. This is of course not the case [since the differential equations do \emph{not} hold charge to be fixed]; the charges are lost, but at a rate too slow to be noticeable at the scale of the plot.   

This behavior comes at a surprise since based on flat space intuition one would expect that a black hole tends to lose charge faster than it loses mass, due to the fact that electromagnetic interaction is so much stronger than gravitational interaction. More precisely, we expect that discharge can only be avoided if gravitational attraction far exceeds Coulomb repulsion for the lightest charged particle pair, namely the electron and positron. This means $Mm/r^2 \gg Qe/r^2$, where $m$ and $e$ are the mass and charge of the electron\footnote{Here, and until further notice, we follow the units and conventions of Hiscock and Weems \cite{kn:HW}, with $G=c=1$ but $\hbar\neq 1$. Consequently, $\hbar G/c^3 = \hbar \approx 2.61 \times 10^{-66} ~\text{cm}^2$. Also, Boltzmann constant $k_B=1$. Without loss of generality, we will choose the charge of the black hole to be positive. The unit of charge follows the Gaussian system, and is such that $M=Q$ defines the extremal [asymptotically flat] Reissner-Nordstr\"om geometry.}. A black hole is thus expected to discharge down to $Q/M \ll m/e \approx 10^{-21}$. In fact, Hiscock and Weems \cite{kn:HW} [see also \cite{kn:SP}] investigate sufficiently massive charged black holes in asymptotically flat spacetime and showed that this is indeed what happens --- the black holes always \emph{eventually} tend towards the Schwarzschild limit, although the charge-to-mass ratio is not necessarily monotonically decreasing depending on the exact initial conditions\footnote{This behavior is not universal for all models of Hawking radiation, see e.g., \cite{wenyu}, in which asymptotically flat charged black holes have different final fate due to different physical assumptions made.}.

This raises an interesting question: why does charged flat black holes behave so much differently than their asymptotically flat counterparts\footnote{The fact that Schwinger process in AdS is \emph{less} efficient than asymptotically flat space has been observed before in the literature \cite{CM, KP, PJ}.  Intuitively, \emph{positive} cosmological constant, e.g., in de-Sitter cosmology, helps to push particle pairs apart as space expands and thus enhances Schwinger effect, whereas \emph{negative} cosmological constant suppresses Schwinger effect.}? This puzzle is even more pronounce if one considers the fact that the work in \cite{omc} concerns black hole spacetimes with \emph{large} AdS length scale $L \gg 10^8$ cm, i.e., \emph{small cosmological constant} $\Lambda = -3/L^2 < 0$, and it does not seem obvious why an asymptotically flat spacetime with $\Lambda = 0$ allows charge loss to be so much more effective than asymptotically AdS one with $|\Lambda| \approx 0$. In this work, we clarify the underlying physics of charge loss, starting with a more detailed analysis of the original work of Hiscock and Weems concerning asymptotically flat charged black holes, and then moving on to the case of AdS black holes.

\section* {\large{\textsf{2. Evolution of Asymptotically Flat Reissner-Nordstr\"om Black Holes}}}

The metric of an asymptotically flat Reissner-Nordstr\"om black hole is
\begin{equation}
g[\text{AFRN}] = -\left(1-\frac{2M}{r}+\frac{Q^2}{r^2}\right)dt^2 + \left(1-\frac{2M}{r}+\frac{Q^2}{r^2}\right)^{-1}dr^2 + r^2 \left(d\theta^2 + \sin^2\theta d\phi^2 \right).
\end{equation}
The horizon of the black hole is located at coordinate radius
\begin{equation}
r_h = M + \sqrt{M^2 - Q^2},
\end{equation}
and the Hawking temperature is given by
\begin{equation}
T = \frac{\hbar\sqrt{M^2-Q^2}}{2\pi (M + \sqrt{M^2 - Q^2})^2}.
\end{equation}

Due to the difficulty in modeling evaporation of charged black holes in general, Hiscock and Weems [henceforth, HW] restricted their investigation to sufficiently large black holes. Since the temperature is inversely proportional to radius for these black holes, they are also necessarily very cold. Hawking \cite{Hawking2} calculated that the number of particles of the $j$th species with charge $e$ emitted in a wave mode labeled by frequency $\omega$, spheroidal harmonic $l$, and helicity $p$ is given by [if we ignore angular momentum of emitted particles and rotation of the hole]
\begin{equation}\label{Hawking}
\left\langle N_{j\omega l p} \right\rangle = \frac{ \Gamma_{j\omega l p}}{\exp\left((\omega-e\Phi)/T\right) \pm 1}, 
\end{equation}
where $T$ is the temperature of the black hole. The plus sign in the denominator corresponds to fermion, while minus sign corresponds to boson. Here $ \Gamma_{j\omega l p}$ denotes the absorption probability for an incoming wave of the specific mode.

In the low temperature case, Gibbons \cite{Gibbons} showed that emission rate of charged particles is well described by Schwinger process \cite{Schwinger}, while thermal emission is considered to only produce \emph{massless} particles. It is important to note, as HW did, that although charged particle emission can be modeled separately from the thermal Hawking flux of neutral particles, they are all part of Hawking radiation. This is because emission of charged particle is thermodynamically related to a chemical potential associated with the electromagnetic field of the black hole. Furthermore, from Eq.(\ref{Hawking}) we see that the precise statement is actually the following: At \emph{all} nonzero temperature $T$, all species of particles regardless of whether they are charged, \emph{are} emitted by Hawking radiation. However, at low temperature, production of charged [and therefore massive] particles is \emph{exponentially suppressed} by the Boltzmann factor. In the model adopted by HW, this suppression, as we will see, is realized via the exponential term in the Schwinger process\footnote{The model has limitations. For example, Schwinger emission is of course \emph{not} thermal.}, which describes the rate of pair creation per unit 4-volume $\Gamma$ by
\begin{equation}
\Gamma = \frac{e^2}{4\pi^3 \hbar^2}\frac{Q^2}{r^4} \text{exp}\left(-\frac{\pi^2 m^2 r^2}{\hbar e Q}\right) \times \left[1 + O\left(\frac{e^3Q}{m^2 r^2}\right)\right].
\end{equation}
A characteristic scale involved in the Schwinger process is the Schwinger critical charge $E_c :=\pi m^2/(\hbar e)$. For convenience, HW denote its inverse by $Q_0 := \hbar e/(\pi m^2)$.

Indeed, in addition to the ``weak-field approximation'' in which one ignores all higher order terms, HW also apply the series approximation\footnote{Note that this series \emph{diverges} for all $x > 0$, but if a fixed number of terms is taken, then for large enough $x$, the approximation is good. However, the divergence means that, for any fixed $x$, increasing the number of terms in the series does not help to increase the accuracy of the approximation. Such series is called an \emph{asymptotic series}.} for the complementary error function $\text{erfc}(x) = 1- \text{erf}(x)$, namely,
\begin{equation}\label{series}
\text{erfc}(x) = \frac{e^{-x^2}}{x\sqrt{\pi}}\left[1 + \sum_{n=1}^{\infty}  (-1)^n \frac{1 \cdot 3 \cdot 5 \cdots (2n-1)}{(2x^2)^n} \right], ~~x \gg 1,
\end{equation}
to the charge loss rate [obtained from integrating $\Gamma$]
\begin{flalign}
\frac{dQ}{dt} &\approx \frac{e^3}{\hbar^2} \int_{r_h}^\infty \frac{Q^2}{r^2} \exp{\left(-\frac{r^2}{Q_0Q}\right)} dr\\
& =\frac{e^3}{\pi^{2} \hbar^2} \left[-\frac{Q^{3/2}\sqrt{2}}{\sqrt{Q_0}} ~\text{erf}\left(\frac{r}{\sqrt{Q_0 Q}} \right)- \frac{Q^2}{r} \exp{\left(-\frac{r^2}{Q_0 Q}\right)}\right]\Bigg|_{r_h}^\infty.
\end{flalign}
Thus, they obtained, finally, the ordinary differential equation that governs charge loss as
\begin{equation}\label{chargeloss0}
\frac{dQ}{dt} \approx - \frac{e^4}{2\pi^{3} \hbar m^2} \frac{Q^3}{r_h^3} \exp{\left(-\frac{r_h^2}{Q_0Q}\right)}.
\end{equation}

Note that the series approximation applying to $\text{erfc}(r_h/\sqrt{QQ_0})$ means that HW are necessarily only restricting their analysis to the case
\begin{equation}
r_h^2 \gg QQ_0 \Longleftrightarrow \frac{Q}{r_h^2} \ll E_c. 
\end{equation}
That is to say, charged particle production is greatly suppressed as required. Thus, despite the occurrence of Schwinger formula, for the model to be self-consistent, we actually want the Schwinger effect [which produces copious amount of charged particles] to \emph{not} set in. In other words, \emph{charge loss is inefficient} in the regime of validity of the model. Therefore the puzzle is more appropriately phrased as follows: why is charge loss \emph{so much more} inefficient in the case of charged flat black holes than their asymptotically flat cousins? 

With this question in mind, let us first review the results of Hiscock and Weems. Having introduced the physics of charge loss, 
the mass loss of the black hole can be described by
\begin{equation}\label{massloss0}
\frac{dM}{dt} = -\alpha aT^4\sigma + \frac{Q}{r_h}\frac{dQ}{dt}.
\end{equation}
The first term on the right describes thermal mass loss due to Hawking radiation; which is just the Stefan-Boltzmann law, with $a=\pi^2/(15\hbar^3)$ denoting the radiation constant\footnote{This is $4/c$ times the Stefan-Boltzmann constant, although HW refer to $a$ as simply the ``Stefan-Boltzmann constant''}. 
The quantity $\sigma$ denotes the area of the emitting surface, which is \emph{not} the event horizon but the surface that corresponds to the [unstable] photon orbit. The reason is that only particles that have enough energy can escape the effective potential barrier, with local maximum at the photon orbit [see Fig.(6.5) of \cite{Wald}].
The constant $\alpha$ depends on the number of species of massless particles; it is essentially the so-called ``grey-body factor''. Due to the huge time scale involved in the life time of black holes, and the fact that $\alpha$ only gives order one correction \cite{kn:HW}, we will henceforth set $\alpha=1$ for simplicity. The second term in Eq.(\ref{massloss0}) is of course due to mass loss of charged particles. It is in fact the same term that appears in the first law of black hole mechanics in general relativity: $dM=(\kappa/8\pi) dA + (Q/r_h) dQ + \Omega dJ$. 

Eq.(\ref{chargeloss0}) and Eq.(\ref{massloss0}) form a system of coupled linear ODEs, which can be numerically solved once the initial mass and initial charge are specified.
HW found that although asymptotically flat charged black holes always evolve towards the Schwarzschild limit, the evolutionary path each black hole takes depends on the initial charge-to-mass ratio $Q/M$. For low $Q/M$ ratio, the black holes are in ``mass dissipation zone'' --- they lose mass faster than charge and thus actually, \emph{initially} tend towards extremal limit. Their specific heat changes sign from negative to positive. Eventually however, their evolution leaves the positive specific heat region of the parameter space, and they flow along an attractor that brings them towards the Schwarzschild limit. 

\begin{figure}[!h]
\centering
\mbox{\subfigure{\includegraphics[width=3in]{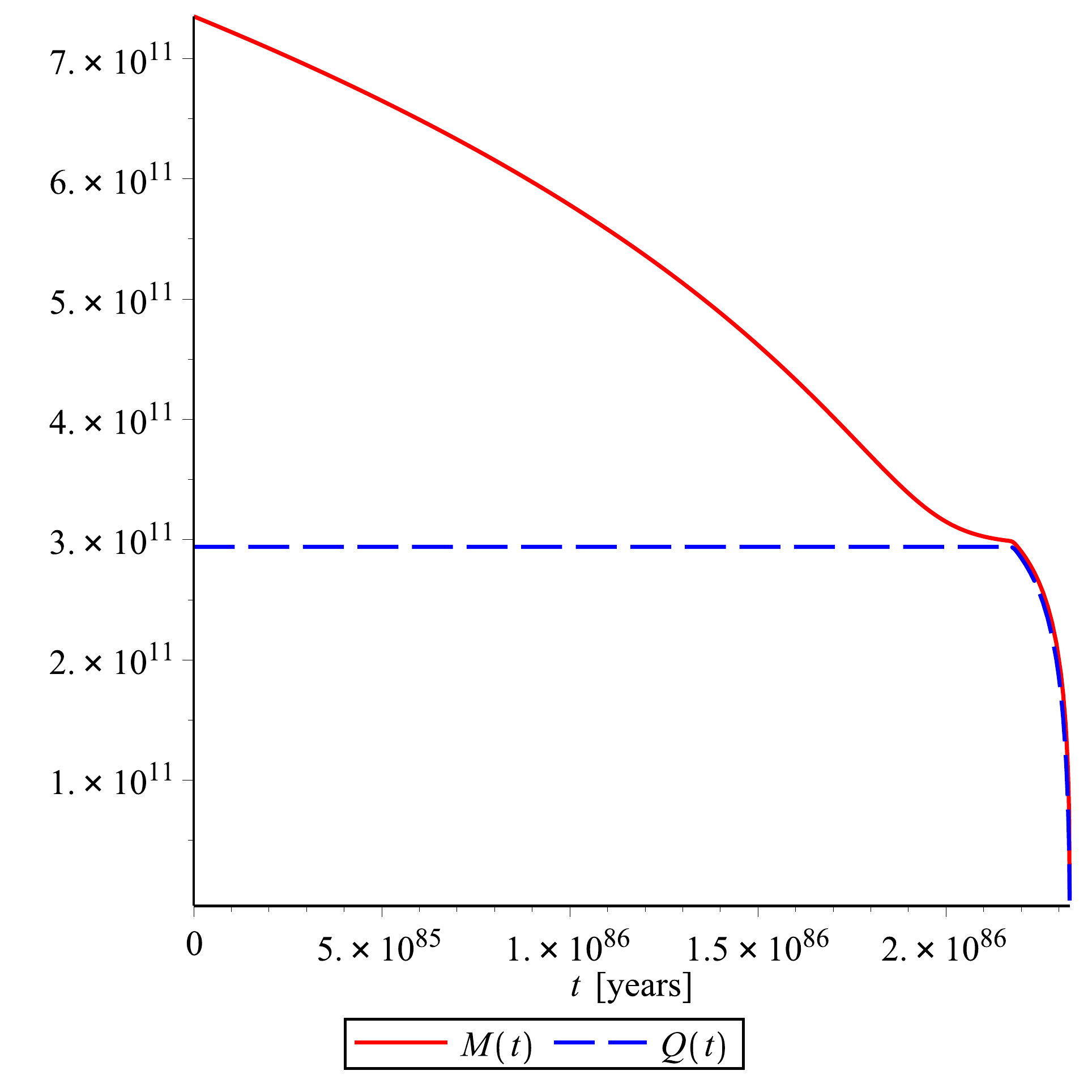}}\quad
\subfigure{\includegraphics[width=3in]{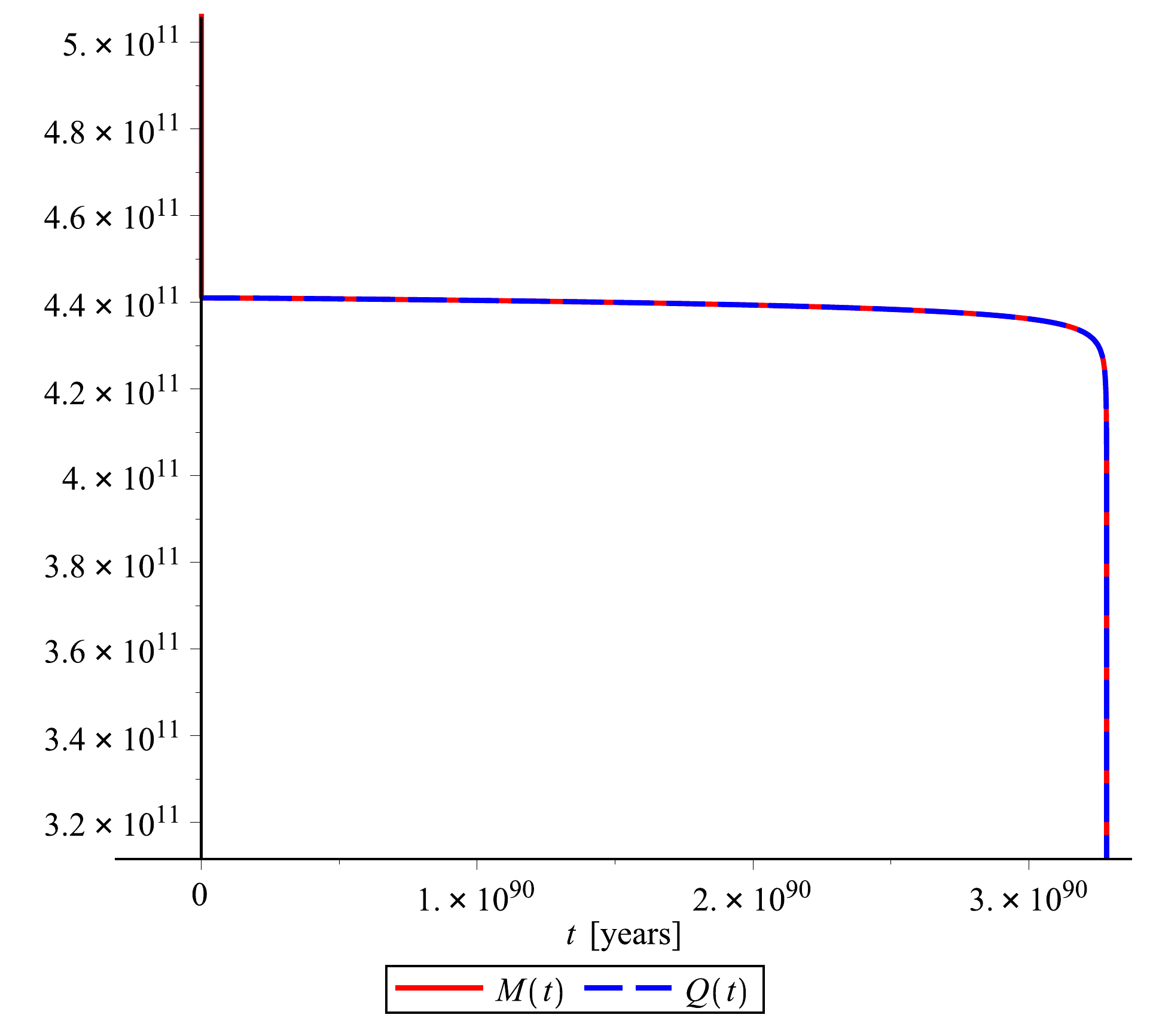} }}
\caption{The evolution of mass and charge of asymptotically flat Reissner-Nordstr\"om black hole in the mass dissipation zone. The initial conditions are $M(0)=7.35 \times 10^{11}$ cm, $Q(0)=2.94 \times 10^{11}$ cm for the left figure, and  $M(0)=7.35 \times 10^{11}$ cm, $Q(0)=4.41 \times 10^{11}$ cm for the right figure. \label{2}}
\end{figure}

One interesting feature for these black holes is that, while electrical charge stays almost constant initially, mass steadily decreases, until $M \sim Q$, and then they start to evolve together [since for $M \sim Q$, we have $T \sim 0$ and $dM/dt \sim dQ/dt$] for some time. Consequently the black hole -- depending on the exact initial conditions [such as the one in the right plot of Fig.(\ref{2})] -- can stay near extremal limit for a long time, until $Q/M$ starts to decrease. 

\begin{figure}[!h]
\centering
\mbox{\subfigure{\includegraphics[width=3.2in]{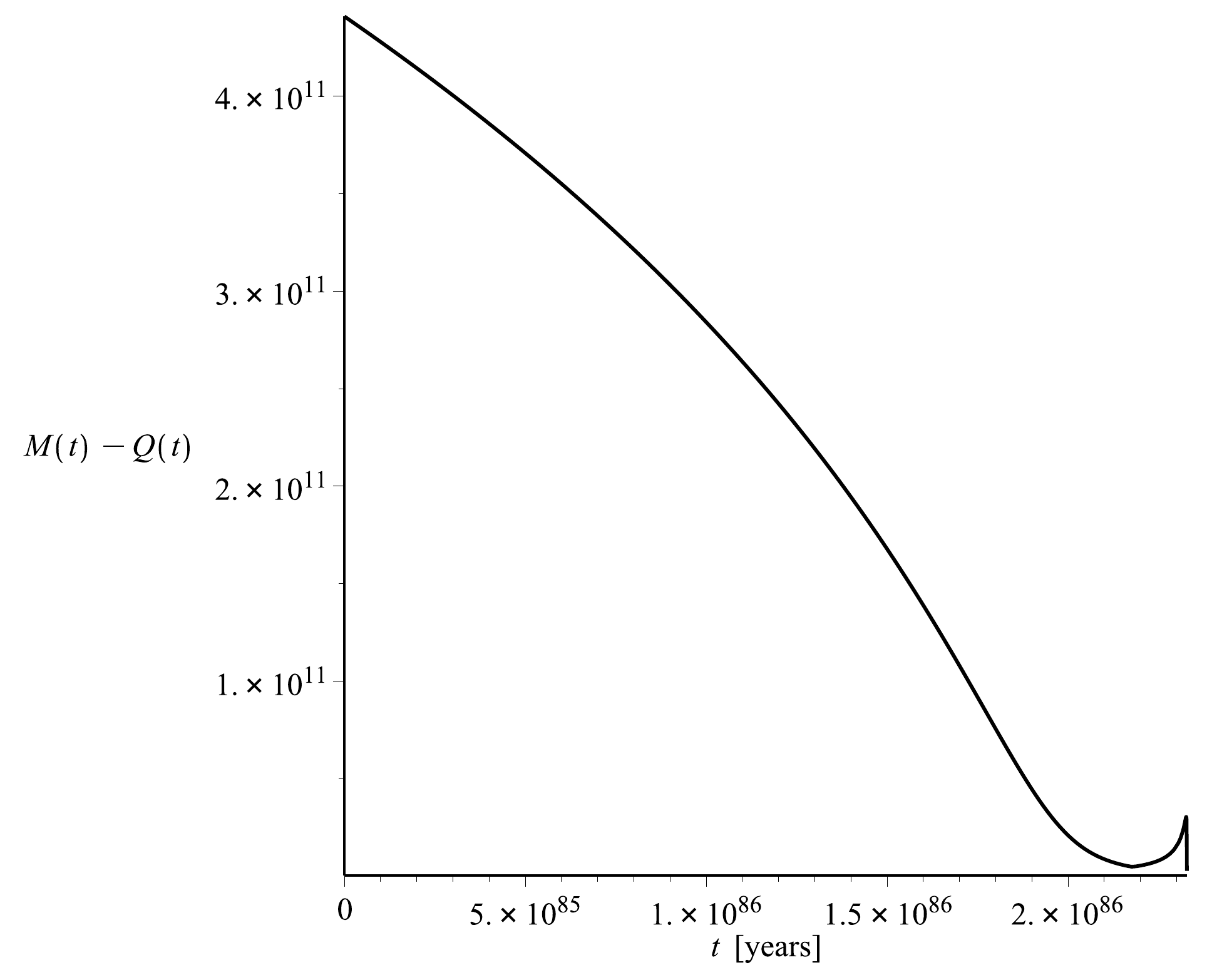}}\quad
\subfigure{\includegraphics[width=3in]{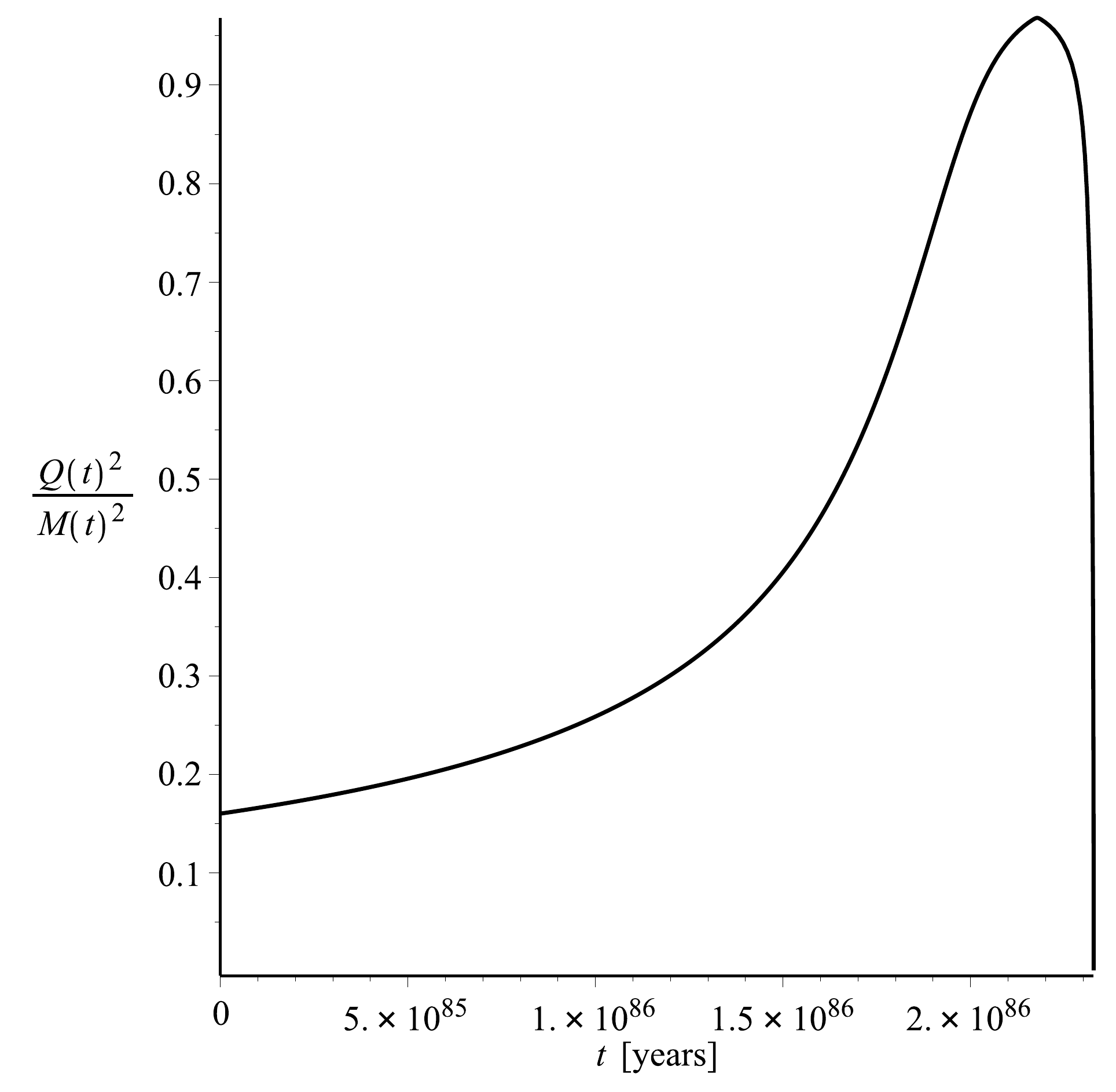} }}
\caption{\textbf{Left:} The evolution of mass-charge difference, $M-Q$, of an asymptotically flat Reissner-Nordstr\"om black hole in the mass dissipation zone. \textbf{Right:} The evolution of charge-to-mass ratio of the same black hole. In this example the initial conditions are $M(0)=7.35 \times 10^{11}$ cm and $Q(0)=2.94 \times 10^{11}$ cm. Note that \emph{initially} the charge-to-mass ratio increases, but eventually decreases towards Schwarzschild limit.\label{3}}
\end{figure}

Note that although it may appear from the plots that $Q \sim M$ even at this stage, this is only because the scale does not resolve the two curves close enough to see the difference between them. For easier comparison we can plot both $M-Q$ as a function of time [see Fig.(\ref{3})] in which it is evident that the difference between $M$ and $Q$ can be large towards the end. Note also that the eventual decrease in $M-Q$ is not inconsistent with the decrease in $Q/M$. After all, $d(M-Q)/dt \neq d(Q/M)/dt$. Indeed, we see that
\begin{equation}
\frac{d}{dt}\left(\frac{Q}{M}\right)=\frac{1}{M}\left[\frac{dQ}{dt}-\frac{Q}{M}\frac{dM}{dt}\right]
\end{equation}  
can be negative, i.e., $Q/M$ is decreasing, if 
\begin{equation}
\frac{dQ}{dt} < \frac{Q}{M}\frac{dM}{dt}.
\end{equation}
Recall that $dM/dt$ and $dQ/dt$ are both negative. Thus this is equivalent to
\begin{equation}
\left|\frac{dQ}{dt}\right| > \frac{Q}{M}\left|\frac{dM}{dt}\right|.
\end{equation}
Therefore $d(Q/M)/dt$ can be negative even if $dM/dt < dQ/dt$, or equivalently, $|dM/dt| > |dQ/dt|$, provided that $Q/M$ is small enough. 
This is precisely what happens towards the end of the evolution as depicted in Fig.(\ref{3}). 

\begin{figure}[!h]
\centering
\mbox{\subfigure{\includegraphics[width=3.0in]{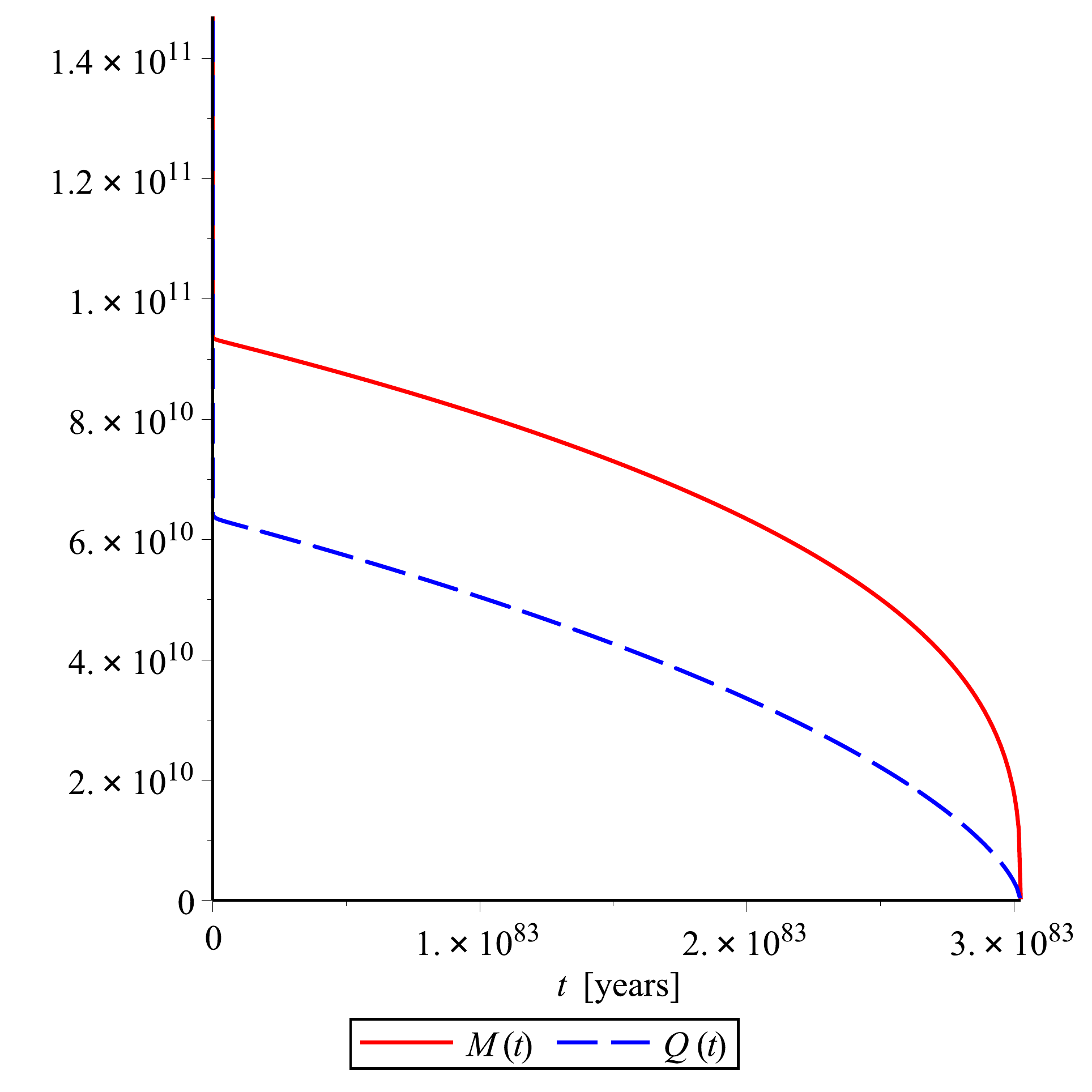}}\quad
\subfigure{\includegraphics[width=3.0in]{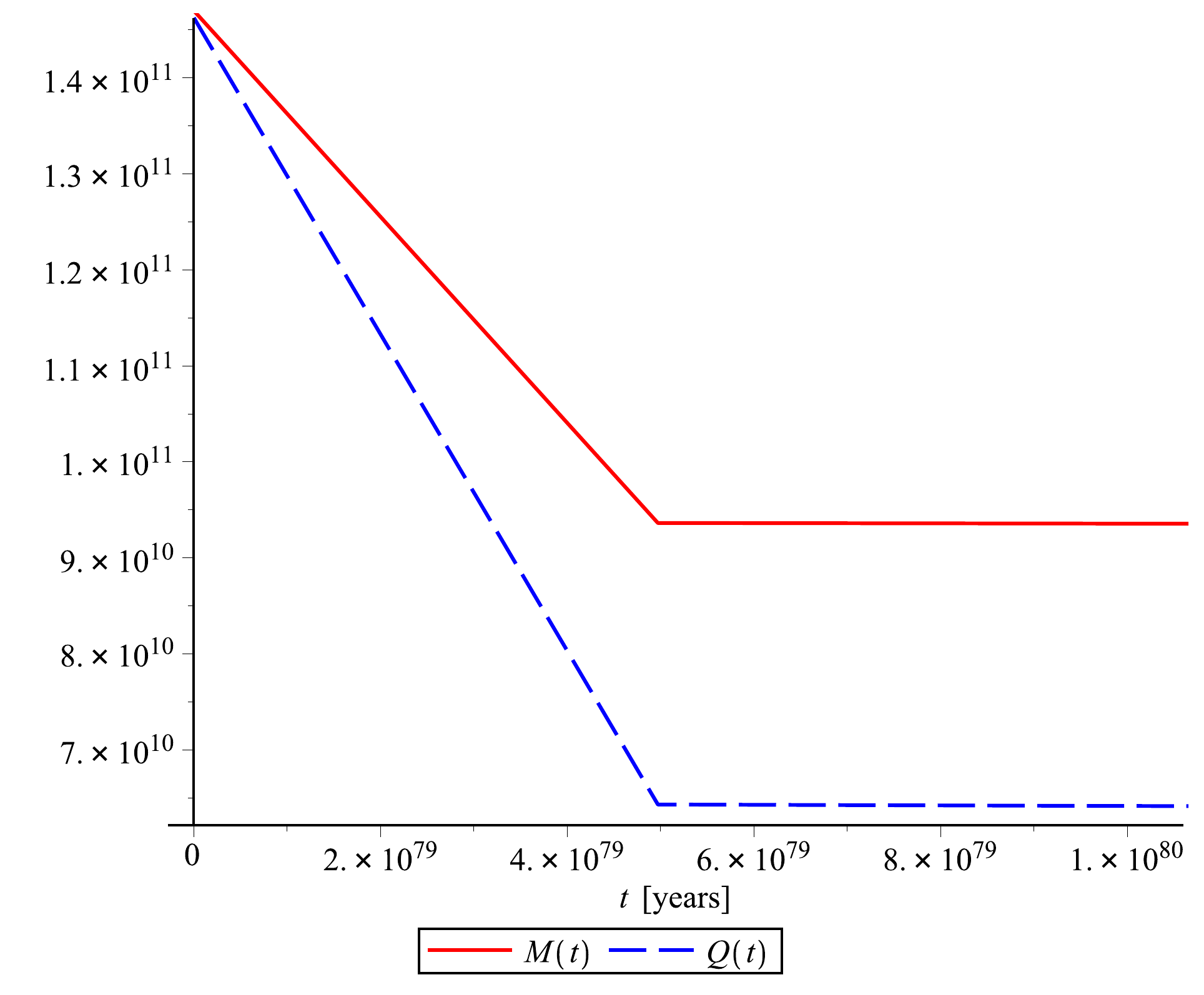} }}
\caption{\textbf{Left:} The evolution of charge-to-mass ratio of a highly charged [i.e., in the charge-dissipation zone] black hole with initial conditions $M(0)=1.47 \times 10^{11}$ cm and $Q(0)^2/M(0)^2=0.99$. \textbf{Right:} Part of the same plot now enlarged to show that the initial ``rapid'' drop of charge and mass actually spans over $O(10^{79})$ years. \label{4}}
\end{figure}

Highly charged black holes however, are in ``charge dissipation zone'' -- they lose their charge steadily and evolve towards Schwarzschild limit without any surprising behavior \cite{kn:HW}. Despite the fact that it looks like both charge and mass drop rapidly when one plots the entire evolution of the black hole [see the left plot of Fig.(\ref{4})], this is again an illusion due to the scale involved. If one zooms in to the ``rapid drop'' portion of the graph, it becomes clear that the process takes quite a long time by ``normal'' standard [although short relative to the much longer time required to decode Hawking radiation], specifically, $O(10^{79})$ years in the example plotted [see the right plot of Fig.(\ref{4})]. This is consistent with the fact that charge loss is \emph{not} supposed to be rapid. 

\begin{figure}[!h]
\centering
\includegraphics[width=0.90\textwidth]{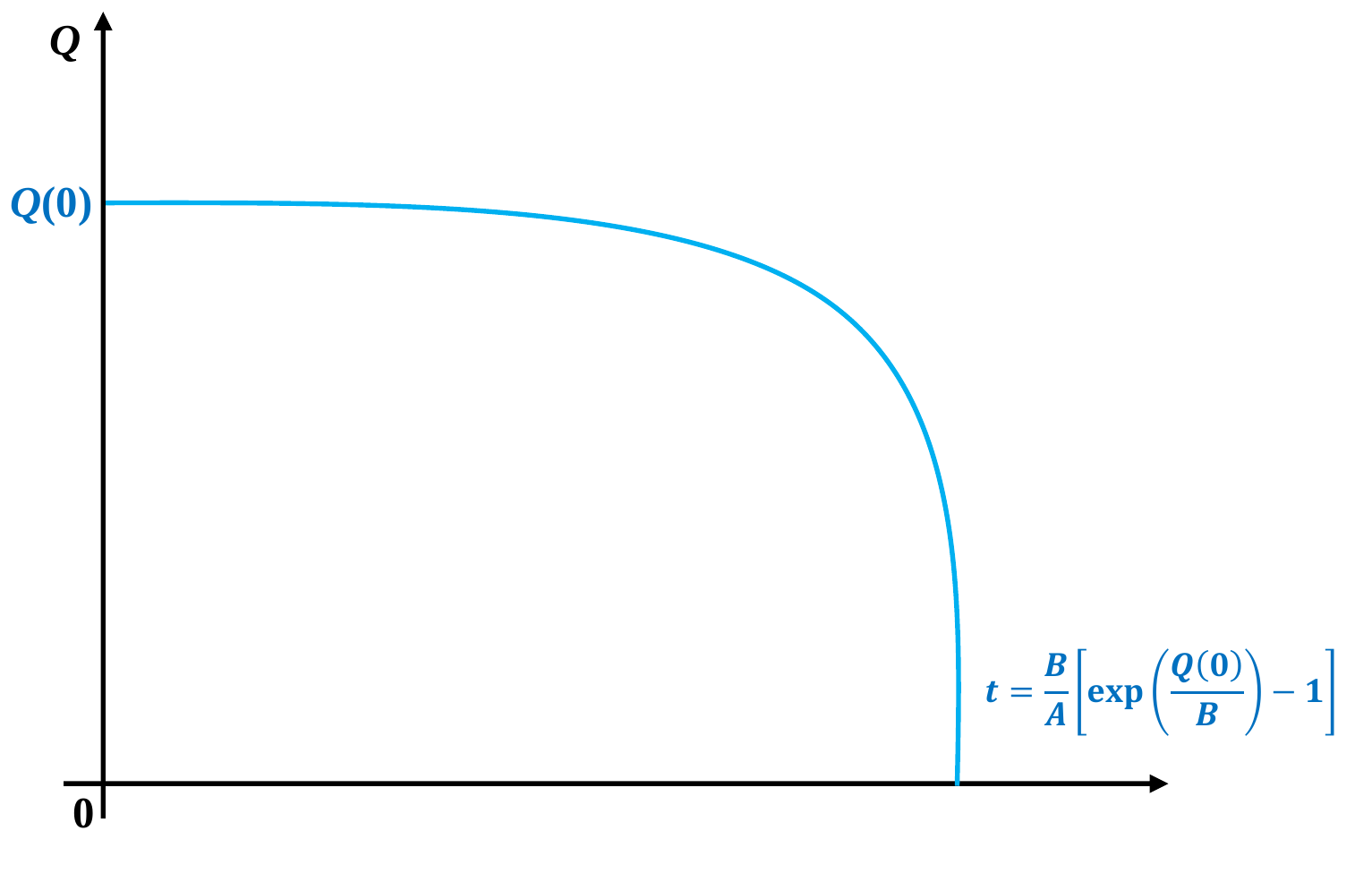}
\caption{The evolution of charge for a generic extremal black hole. Here $A=-e^4/(2\pi^3\hbar m^2)$ and $B=Q_0=\hbar e/(\pi m^2)$. \label{5}} 
\end{figure}

At this point of the discussion, it is insightful to consider an \emph{extremal} black hole, characterized by $M=Q$. Its horizon is located at $r_h=M$. Note that an extremal black hole has absolute zero temperature; it does not emit any Hawking radiation\footnote{There is a large literature on whether semi-classical extremal black hole exists [see e.g., \cite{AHT1, Lowe, AHT2}]; even at the classical level it is not clear what would the final state of an extremal black hole be since it appears to be unstable \cite{Aretakis1, Aretakis2, Aretakis3, LMRT, MRT, ZWA}. Here we are neither concerned about the actual physical existence nor the stability of such solution --- we are merely interested in the mathematical solution as it provides insight into the more complicated \emph{non-extremal} case.}. Nevertheless, the ODE system of Hiscock and Weems still works --- it reduces to only one ODE governing the charge loss rate:
\begin{equation}
\frac{dQ}{dt} \approx -\frac{e^4}{2\pi^3\hbar m^2} \exp\left(-\frac{\pi m^2 Q}{\hbar e}\right).
\end{equation}
Of course this ODE \emph{ceases to model Hawking radiation}, however it nevertheless still describes charge-loss of extremal black hole via [non-thermal] Schwinger process \cite{CM, Khriplovich, Khriplovich2}. This ODE has the form
\begin{equation}
\frac{dQ}{dt} = -A \exp\left(-\frac{Q}{B}\right),
\end{equation}
which is readily solved to yield [see Fig.(\ref{5})].
\begin{equation}
Q(t) = B \left[\ln\left(\exp\left(\frac{Q(0)}{B}\right)-\frac{At}B\right)\right].
\end{equation}
The function $Q(t)$ stays more or less constant initially but then eventually starts to drop and becomes zero at
\begin{equation}
t = A^{-1}B \left[\exp\left(\frac{Q(0)}{B}\right)-1\right].
\end{equation}

One can see that this behavior is essentially the same as how charge evolves for \emph{non-}extremal black holes in the mass dissipation zone, although the time it takes for $Q(t)$ to vanish is extended some what. This is understandable since with thermal correction charge dissipation becomes slower [near-extremal hole has larger surface area and thus smaller electric field than exactly extremal hole]. One can check numerically that the lifetime of charged black holes, regardless of whether they started off in the mass dissipation zone or the charge dissipation zone, is always longer than an extremal hole of the same initial charge.   

We now turn our attention to charged flat black holes.

\section* {\large{\textsf{3. The Curious Case of Charged AdS Flat Black Holes}}}

In \cite{omc}, the analysis of Hiscock and Weems was extended to asymptotically locally AdS charged black holes with flat horizon. 
In four-dimensions, such charged flat black holes have metrics of the form [see \cite{kn:77}]
\begin{equation}\label{ARRGH}
g[\text{AdSRN}(k=0)] = -\, \Bigg[{r^2\over L^2}\;-\;{8 \pi M^*\over r}+{4\pi Q^{*2}\over r^2}\Bigg]dt^2\; + \;{dr^2\over {r^2\over L^2}\;-\;{8 \pi M^*\over r}+{4\pi Q^{*2}\over r^2}} \;+\;r^2\Big[d\zeta^2\;+\;d\xi^2\Big],
\end{equation}
where $\zeta$ and $\xi$ are dimensionless coordinates on a flat 2-dimensional Riemannian manifold [and thus has scalar curvature $k=0$]. $M^*$ and $Q^*$ are mass and charge parameters, respectively. They are defined as follows. If the horizon is compact, we take it to be a flat square torus with area 4$\pi^2 K^2$, in which $K$ is a dimensionless ``compactification parameter'', and define $M^*$ as $M/(4\pi^2 K^2)$. Similarly we define $Q^* = Q/(4\pi^2 K^2)$. Here $M$ and $Q$ are the physical mass and charge of the hole. [In AdS/CFT application, the mass and charge densities of the black hole in the AdS bulk are \emph{defined} by the field theory on the boundary]. In the case of planar [non-compact] horizon, we simply let $M$, $Q$, and $K$ tend to infinity in such a way that the quantities $M^*$ and $Q^*$ always remain finite. Following the convention in \cite{omc}, here we employ Heaviside-Lorentz unit for the electric charge in the case of flat charge black hole, so that instead of $Q$ in Gaussian unit, we now write $Q/\sqrt{4\pi}$.

It can be checked that, given any mass $M$, the extremal charge is 
\begin{equation}
Q_{\text{ext}}= (108 \pi^5 M^4 L^2 K^4)^{1/6}.
\end{equation}
It is therefore convenient to define a dimensionless function 
\begin{equation}\label{w}
w[M]:=\frac{(108\pi^5L^2K^4)^{1/6}}{M^{1/3}},
\end{equation}
so that the [normalized] charge-to-mass ratio becomes unity at extremality. Indeed, we have
\begin{equation}\label{QwM}
\frac{Q}{wM} \in [0,1].
\end{equation}

The ODE system is then obtained in more or less the same manner as HW did. Nevertheless, there are many subtleties involved. For example, the area of the radiative surface is independent on the mass of the black hole, and is simply $4\pi^2 K^2 L^2$. Furthermore, the validity of series approximation for Schwinger formula is also mass independent. Instead it only requires large $L$, specifically $L \gg 10^8$ cm. See \cite{omc} for details.

\begin{figure}[!h]
\centering
\includegraphics[width=0.70\textwidth]{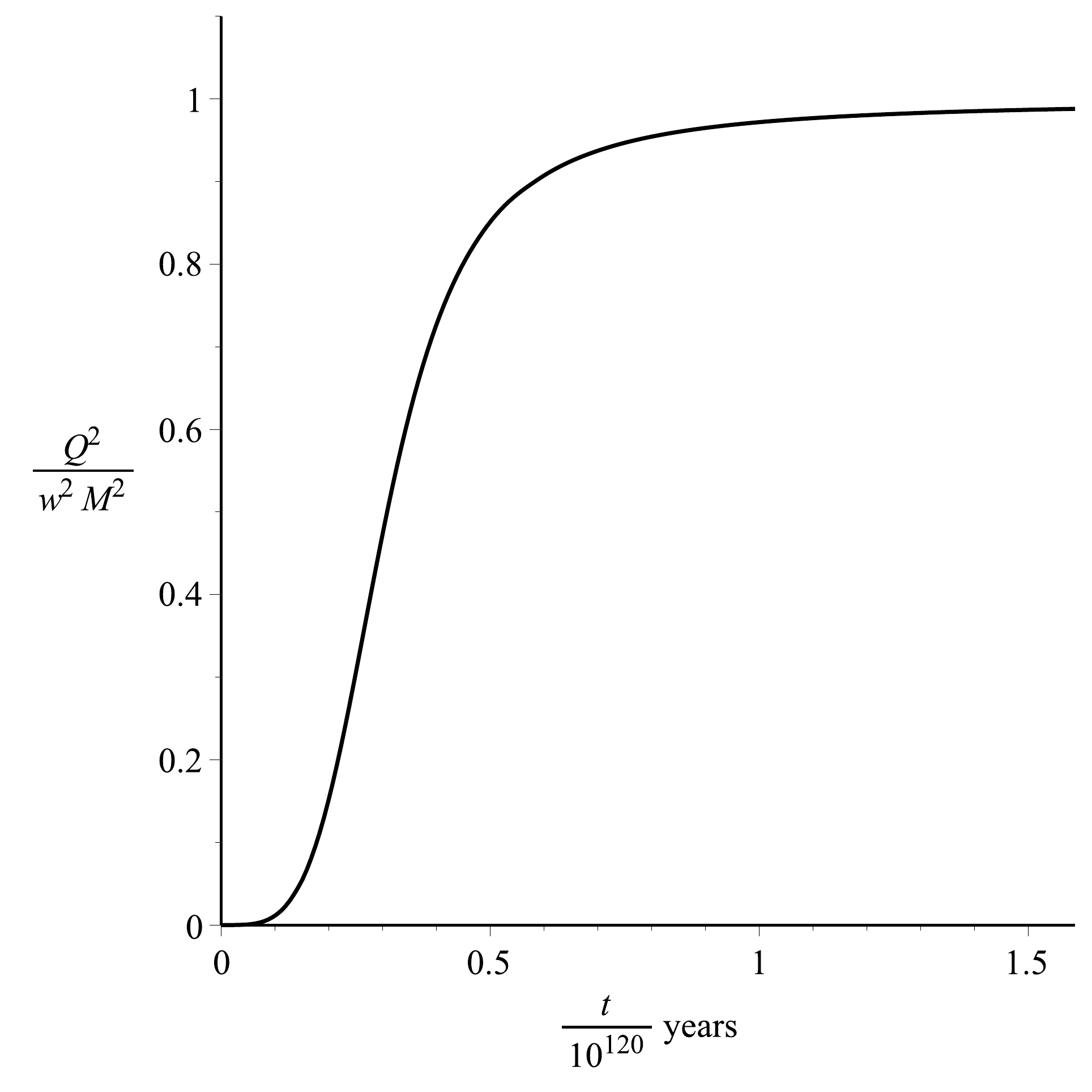}
\caption{The square of the normalized charge-to-mass ratio as a function of time of a charged toral black hole with compactification parameter $K=1$ and $L=10^{15}$ cm, initial mass $M(0)=5.6 \times 10^{20}$ cm, and initial charge $Q(0)=6 \times 10^{-34}$ cm. \label{6}} 
\end{figure}

Here we only write down the final results:
\begin{equation}\label{ODE2}
\begin{cases}
\dfrac{dM^*}{dt} = - \dfrac{a}{4}L^2 T^4 + \dfrac{Q^*}{r_h}\dfrac{dQ^*}{dt},
\\
\\
\dfrac{dQ^*}{dt} \approx - \dfrac{e^4}{64 \pi^{11/2} \hbar m^2} \dfrac{Q^{*3}}{r_h^3} \exp{\left(-\,\dfrac{2\pi^{3/2}m^2r_h^2}{\hbar e Q^*}\right)},
\end{cases}
\end{equation}
with the Hawking temperature given by
\begin{equation}
T = \frac{\hbar}{r_h^2}\left[6M^* - \frac{4Q^{*2}}{r_h} \right] = \hbar\left[\frac{r_h}{\pi L^2} - \frac{2M^*}{r_h^2}\right].
\end{equation}
Unlike asymptotically flat case in which the black holes in mass dissipative zone change the sign of their specific heat as they become highly charged, we found no such behavior in the case of charged flat black holes [at least for large $L$], and their specific heat always stays positive. Recall that in asymptotically flat case the charge-to-mass ratio of a black hole that started off in the mass dissipative zone initially increases but then decreases as it starts to flow down an attractor in the parameter space. HW showed that the attractor is defined by the positive specific heat region in the parameter space. Therefore, the absence of sign change in the specific heat of charged flat black hole is consistent with the numerical evidence that the [normalized] charge-to-mass ratio always increases. [One such example is provided in Fig.(\ref{6}).]

The reason that the [normalized] charge-to-mass ratio \emph{never} decreases is due to the fact that -- as Fig.(\ref{1}) indicates -- the charge of the black hole never decreases appreciably, unlike asymptotically flat black holes [Fig.(\ref{2}) and Fig.(\ref{4})]. Since we are dealing with large $L$ and so presumably close to asymptotically flat case, it is tempting to ask why the results are so different. The reason is this: asymptotically flat Reissner-Nordstr\"om spacetime is \emph{not} the limit of charged flat black holes as we take $L \to \infty$, as one can check from the metric tensors explicitly. In the toral case this is even more obvious since spacetime is foliated by 2-tori instead of 2-spheres, and one cannot take limit to pass from one \emph{topology} to another. In other words, despite the fact that pair production by Schwinger process can be said to be local [in the sense that particles are produced near the field-emitting body], one should not expect that the results from asymptotically flat spacetime to also hold in an asymptotically \emph{locally} AdS spacetime. 

Of course, one would expect that the $L \to \infty$ limit for an asymptotically AdS charged black hole with \emph{spherical} topology to correctly recover the behavior of the asymptotically flat case, which also has spherical topology. Such black hole has metric of the form [switching back to Gaussian unit for electrical charge, for easier comparison with asymptotically flat case and HW's analysis.]
\begin{equation}
g[\text{AdSRN}(k=1)] = -\left(1-\frac{2M}{r}+\frac{Q^2}{r^2}+\frac{r^2}{L^2}\right)dt^2 + \left(1-\frac{2M}{r}+\frac{Q^2}{r^2}+\frac{r^2}{L^2}\right)^{-1}dr^2 + r^2 d\Omega^2,
\end{equation}
where $d\Omega^2$ is the standard metric on the 2-sphere.
The horizon of the black hole is located at coordinate radius
\begin{equation}
r_h = \frac{108^{\frac{1}{6}}}{6} L^{\frac{2}{3}} \left[\left(\sqrt{2L^2+27M^2} + \sqrt{27M^2}\right)^{\frac{1}{3}}-\left(\sqrt{2L^2+27M^2} - \sqrt{27M^2}\right)^{\frac{1}{3}} \right].
\end{equation}
Given any mass $M$, the extremal charge is given by
\begin{equation}
Q^2_{\text{ext}} = \frac{r_h}{2}(3M-r_h).
\end{equation}
The normalized charge-to-mass ratio is then $Q/(wM)$, where
\begin{equation}
w^2:=\frac{r_h}{2M^2} (3M-r_h).
\end{equation}

\begin{figure}[!h]
\centering
\mbox{\subfigure{\includegraphics[width=3in]{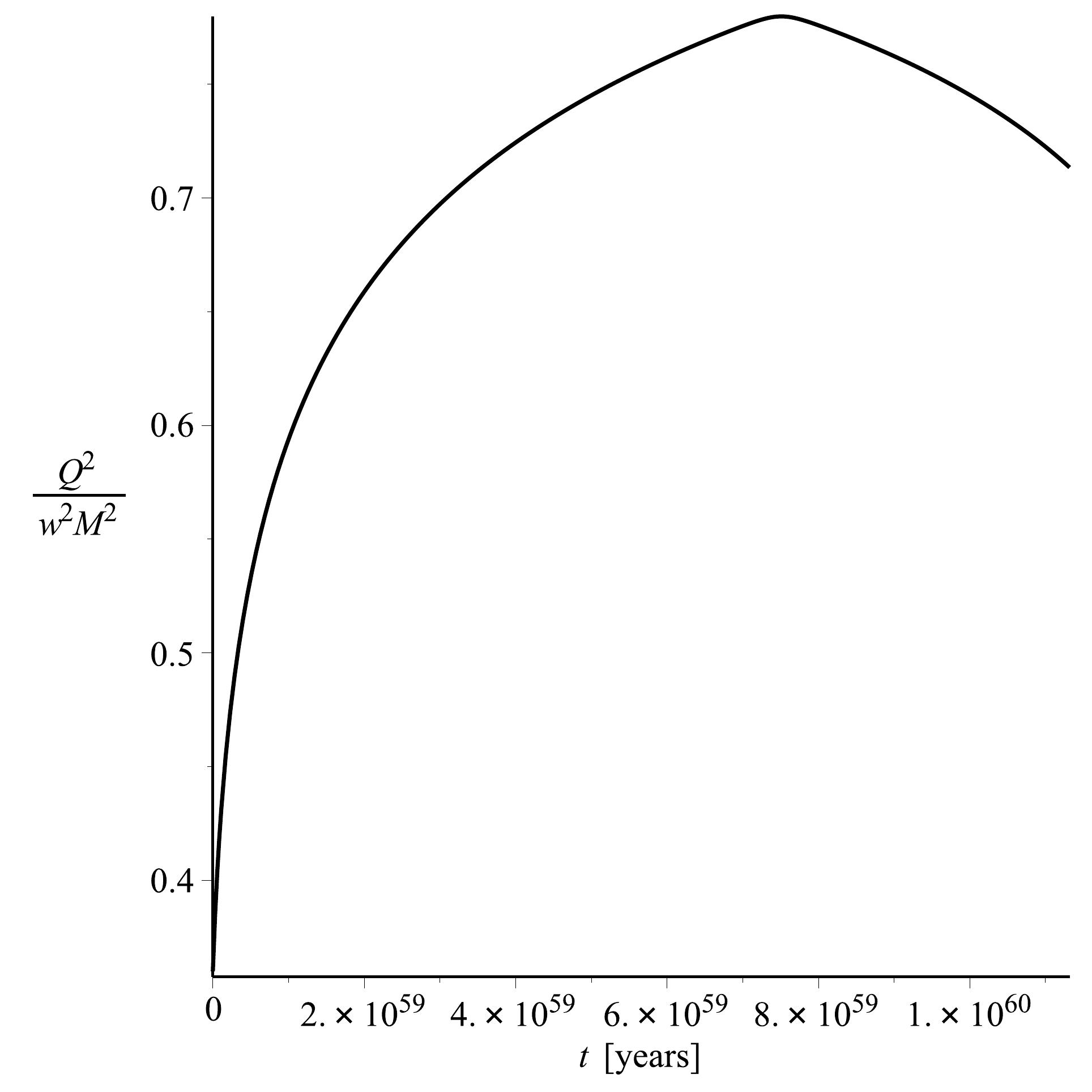}}\quad
\subfigure{\includegraphics[width=3in]{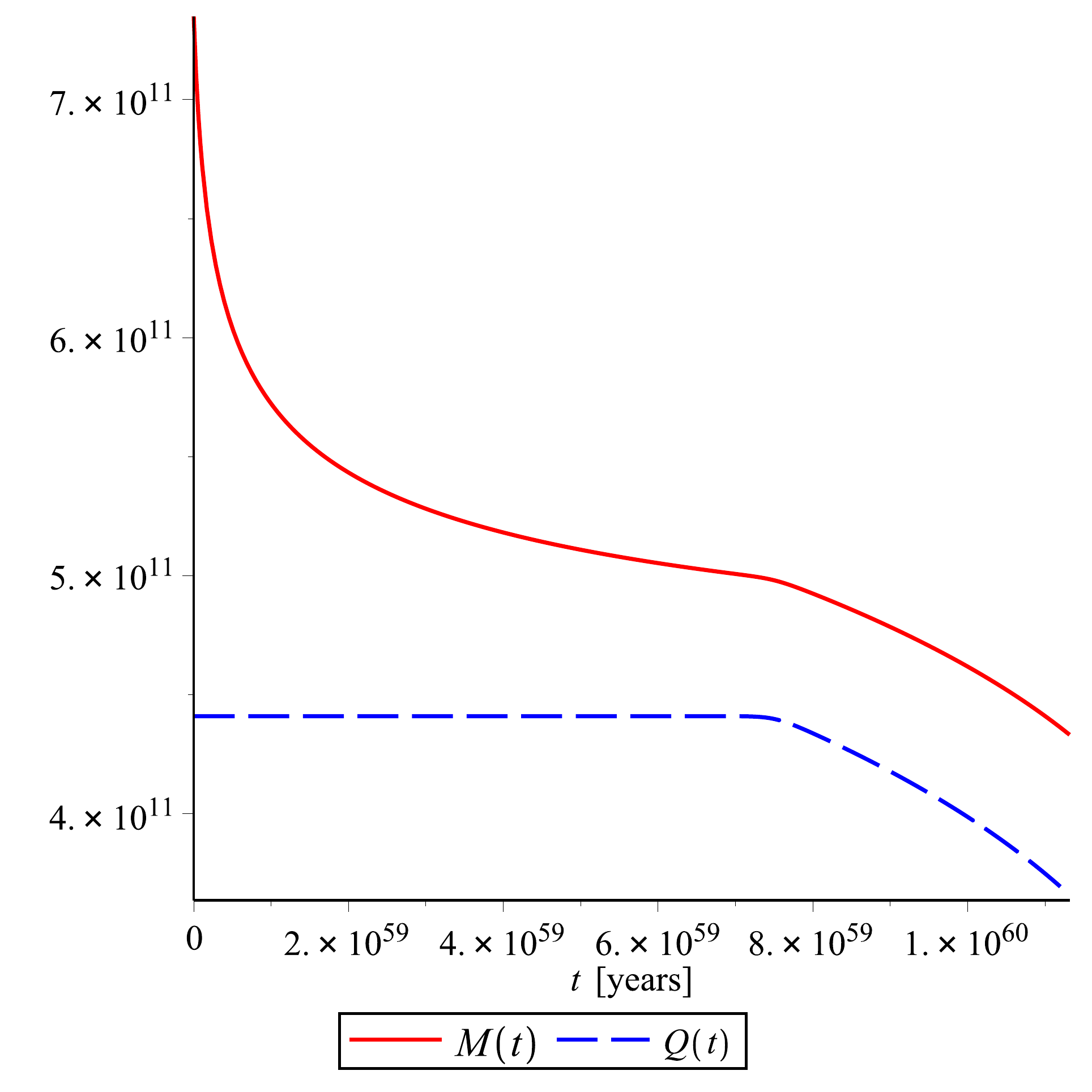} }}
\caption{\textbf{Left:} The evolution of [normalized] charge-to-mass ratio of an AdS charged black hole with spherical topology, with 
initial conditions $M(0)=7.35 \times 10^{11}$ cm and $Q(0)=4.41 \times 10^{11}$ cm. \textbf{Right:} The separate evolutions of mass and charge of the same black hole.
 \label{7}}
\end{figure}

The [unstable] photon orbit in this case \emph{does} depend on $M$ and $Q$, and takes the form \cite{SH, VSOC}
\begin{equation}
r_{\text{ph}} = \frac{3M}{2}\left[1+\sqrt{1-\frac{8Q^2}{9}}\right],
\end{equation}
which reduces to the familiar $r_{\text{ph}}=3M$ value for Schwarzschild black hole when $Q \to 0$. Note also that this expression is independent of $L$. The corresponding impact parameter  $b$ can be calculated straight forwardly, although the expression is complicated and yields no immediate insight to be included here\footnote{In the case of neutral black hole, it is $b^2=\dfrac{27M^2L^2}{27M^2+L^2}$, which reduces to the well-known value $27M^2$ for Schwarzschild geometry when we take $L \to \infty$ limit.}. That expression can then be
substituted into the ODE system of HW, namely Eq.(\ref{massloss0}), with $\sigma = \pi b^2$. The numerical evidence does show the same behavior
as asymptotically flat Reissner-Nordstr\"om black hole, as expected. Namely, 
for black holes which are not highly charged, although their [normalized] charge-to-mass ratio increases at first, that ratio eventually does turn over and tends towards neutral limit [Fig.(\ref{7})].

\begin{figure}[!h]
\centering
\mbox{\subfigure{\includegraphics[width=3in]{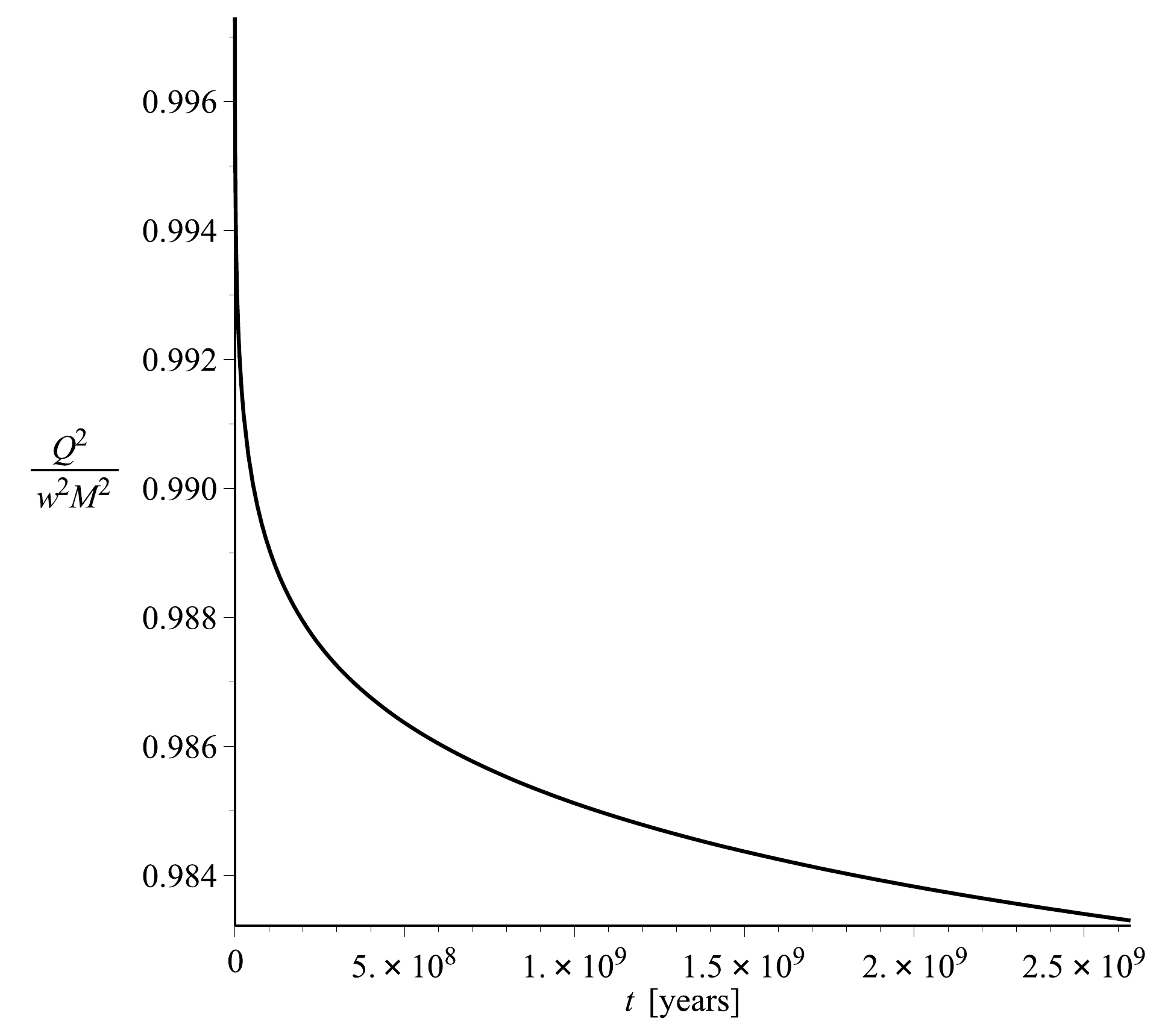}}\quad
\subfigure{\includegraphics[width=3in]{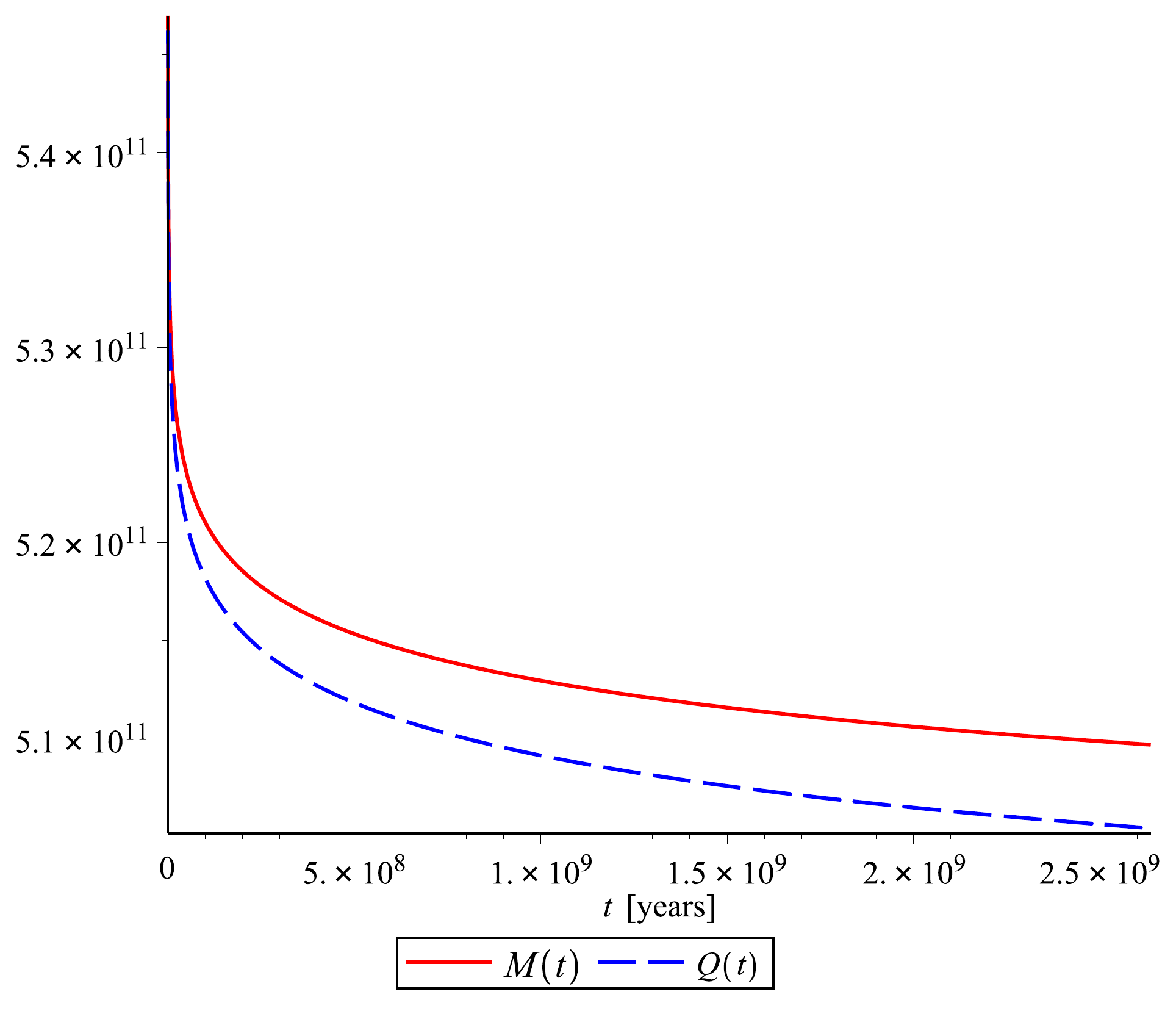}} }
\caption{\textbf{Left:} The evolution of [normalized] charge-to-mass ratio of an AdS charged black hole with spherical topology, with 
initial conditions $M(0)=5.47\times 10^{11}$ cm and $Q(0)=5.462631533\times 10^{11}$ cm. \textbf{Right:} The separate evolutions of mass and charge of the same black hole.  \label{8}}
\end{figure}

\begin{figure}[!h]
\centering
\includegraphics[width=0.80\textwidth]{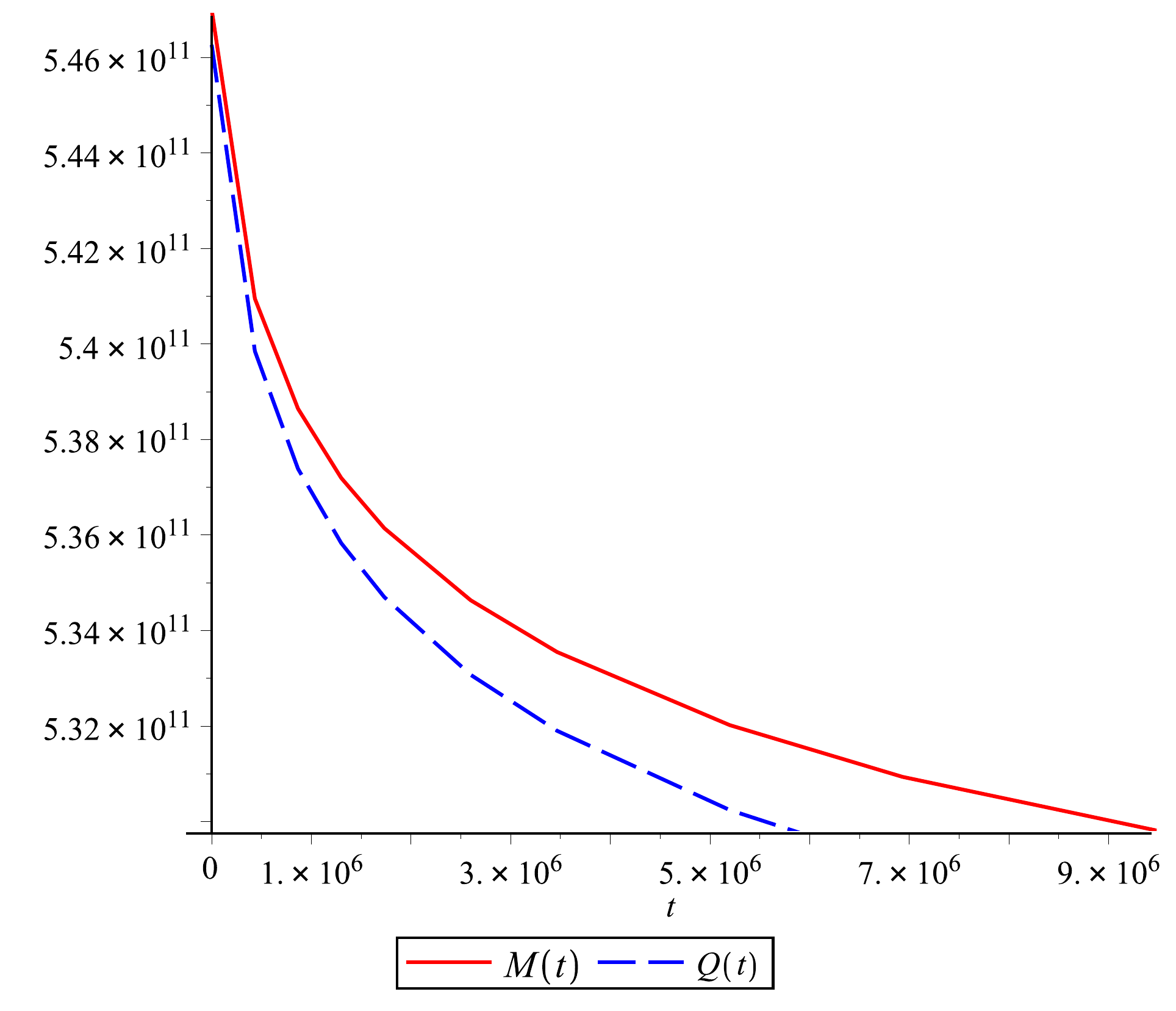}
\caption{The beginning phase of Fig.(\ref{8}) enlarged, showing that the time scale spans across $O(10^6)$ years.
 \label{9}} 
\end{figure}

On the other hand, as expected, the [normalized] charge-to-mass ratio for highly charged black holes simply decreases steadily. Charge loss and mass loss proceed relatively rapidly at the beginning of the evolution [see the right plot of Fig.(\ref{8})], although by ``normal'' standard it takes quite some time as Fig.(\ref{9}) shows.

Having explained why charged flat black holes are not expected to behave like asymptotically flat case even if $L \to \infty$, we still have not explain why charge loss is so much more inefficient for charged flat black holes. To do this we have to take a closer look at the ODE system in Eq.(\ref{ODE2}). Again, let us consider an extremal black hole. As it turns out this already provides an insight into the puzzle since, as we argued before [in the asymptotically flat case], thermal correction for non-extremal black holes only extends the discharge time even more. The objective here is to show that for extremal black hole, the rate of charge loss is practically zero. 

The exponential term in the Schwinger formula is
\begin{equation}\label{Schwinger}
\exp\left(-\frac{2\pi^{3/2}m^2 r_h^2}{\hbar e Q^*_{ext}}\right).
\end{equation}
For charged flat black hole, the extremal horizon is located at
\begin{equation}
r_\text{ext} = \left(\frac{ML^2}{2\pi K^2}\right)^{\frac{1}{3}}=\left(2\pi M^* L^2\right)^{\frac{1}{3}},
\end{equation}
and the extremal charge is given by $Q_{\text{ext}}= (108 \pi^5 M^4 L^2 K^4)^{1/6}$, or equivalently, $Q_{\text{ext}}^*=2^{-1/3}\sqrt{3}L^{1/3}{M^*}^{2/3}\pi^{13/3}$. Upon substituting this into Eq.(\ref{Schwinger}), we find that $M$ and $K$ dependence both drop out of the exponential term, and the charge loss formula now reads
\begin{equation}\label{mainresult}
\frac{dQ^*_{\text{ext}}}{dt} \approx -\frac{AM^*}{L}e^{-BL},
\end{equation}
where
\begin{equation}
A:=\frac{3\sqrt{3}}{2^8}\frac{e^4}{\hbar m^2} \approx 2.4357 \times 10^{39},
\end{equation} 
and
\begin{equation}
B:=\frac{4}{\sqrt{3}}\frac{m^2\pi^2}{\hbar e} \approx 4.5586 \times 10^{-9} ~~\text{cm}^{-1}. 
\end{equation}
Upon evaluating the numbers, one finds that say, if $M \sim L = 10^{15}$ cm, then 
\begin{equation}
\frac{dQ^*_{\text{ext}}}{dt} \sim -O(10^{-1979725}), 
\end{equation}
which is completely negligible. 

The only way for charge loss to become appreciable is to have the black hole mass parameter $M^*$ to be extremely large. However this is not feasible since in our model, following Hiscock and Weems, we need the black hole to be \emph{cold}, so that charge loss can be effectively modeled by Schwinger formula separately from thermal mass loss. This imposes constraint on the mass of the black hole. In \cite{omc}, it was shown that for neutral toral black hole with $K=1$ and $L=10^{15}$ cm, we need $M < O(10^{97})$ cm. Charged black holes can of course tolerate higher bound for mass since its temperature is lower.
Knowing \emph{a posteriori} that the black holes are destroyed when it reaches 92\% of the extremal charge, we can give a somewhat more general bound as follows.

The temperature of the black hole is
\begin{equation}\label{temp1}
T = \frac{\hbar}{2\pi^2K^2}\left[\frac{1}{r_h^2}\left(3M-\frac{Q^2}{2\pi^2 K^2 r_h}\right)\right],
\end{equation}
in which the event horizon can be parametrized by a dimensionless function $\gamma (t) \in [1/2, 2]$ as \cite{omc}
\begin{equation}\label{gamma}
r_h=\left(\frac{\gamma M L^2}{ \pi K^2}\right)^{\frac{1}{3}}=\left(4\pi \gamma M^* L^2\right)^{\frac{1}{3}},
\end{equation}
such that $\gamma=1/2$ corresponds to extremal black hole and $\gamma=2$ to a neutral black hole. At 92\% of the extremal charge, we have $Q/(wM)=0.92$, i.e.
\begin{equation}
\frac{Q^2}{(108 \pi^5 L^2 K^4)^{1/3}M^{4/3}} = 0.92^2.
\end{equation}
This allows us to re-write the expression of Hawking temperature in Eq.(\ref{temp1}) as
\begin{flalign}
T &= \frac{\hbar}{2\pi^2K^2}\left[\frac{1}{r_h^2}\left(3M-\frac{0.92^2\cdot 108^{1/3} M}{2\gamma^{1/3}}\right)\right] \\
&= \left(\frac{\hbar^3 M}{2\pi^{4} \gamma^{2} L^{4} K^{2}}\right)^{\frac{1}{3}} \left[3 - \frac{0.92^2\cdot 108^{1/3}}{2\gamma^{1/3}}\right].
\end{flalign}
This can of course be expressed in terms of $M^*$ and $Q^*$, which would make sense for the planar black hole as well.

Requiring that $T<2m$ yields a bound on $M^*$:
\begin{equation}\label{Mbound}
M^* < \frac{4\pi^2\gamma^2m^3 L^4}{\hbar^3} \left(3-\frac{0.92^2\cdot 108^{1/3}}{2\gamma^{1/3}}\right)^{-3}.
\end{equation}
We see that this bound is quartic in $L$, and therefore has no hope to counteract the effect of the suppression term which is exponential in $L$, for large $L$. Putting in numbers for definiteness by setting $L=10^{15}$ cm, and that $1/2\leq \gamma \leq 2$ taking some value close to $1/2$, we find that $M^* \lesssim O(10^{96})$. This amount -- which is monotonically decreasing -- is then divided by $L$ before multiplying with $10^{-1979725}$, which still yields an extremely small number. 

Of course the rate $dQ_{\text{ext}}^*/dt$ depends on $L$, and indeed upon substituting Eq.(\ref{Mbound}) into Eq.(\ref{mainresult}), we find that $|dQ_{\text{ext}}^*/dt|$
does become order unity for around $L = L_{\text{c}} \sim 1.5 \times 10^{10}$ cm. 
One may thus worry that the black hole may discharge appreciably for low $L \lesssim L_c$. However, recall that our model is only consistent with $L \gg 10^{8}$ cm [more precisely, $L \gg 3.4 \times 10^8$ cm], which as we recall, originated from requiring that Schwinger effect be sufficiently suppressed, and mathematically, from the requirement that the series approximation in Eq.(\ref{series}) holds, which requires $x \gg 1$. The bound $L \gtrsim L_c$ corresponds to $x \gtrsim 10$, which is consistent with $x \gg 1$. [The asymptotic series actually ``converges'' rather quickly as $x$ increases, and so one does not need to go to even higher power to get a good approximation.] 
In other words, the bound $L \gg 10^8$ cm certainly should not be treated as $L \gtrsim 10^8$ cm, but an order or two greater to obtain a good approximation. 
This is the reason the value $L=10^{15}$ cm was used in the numerical work in \cite{omc}, and also the reason why we should not worry that $L \lesssim L_c$ seems to lead to different physics --- this just means that the model already breaks down at that point\footnote{Indeed, a sign that the model breaks down for $L \lesssim L_c$ is that numerical artifacts, e.g. apparent spiking \emph{up} of the charge, start to show up in that range.}, and a separate, careful treatment is needed to model Hawking evaporation. Indeed, for small enough $L$ [though of course still much larger than string length], we expect charge loss to become \emph{efficient}. Nevertheless, mass loss is also more efficient at the same time. Therefore it is not clear that the [normalized] charge-to-mass ratio will evolve differently. We leave this detailed investigation for future work.

Let us now be more explicit in our claim that considering non-extremal black holes does not help to increase charge loss rate. The charge loss for a generic charged hole is similar to Eq.(\ref{mainresult}), but instead of $Q^*_{\text{ext}}=wM^*$, we now have $Q^*=\delta \times (wM^*)$ where $\delta \in [0,1]$, with $\delta=1$ for extremal hole and $\delta = 0$ for neutral hole.
In addition, the horizon is given by Eq.(\ref{gamma}). Therefore one obtains
\begin{equation}\label{mainresult2}
\frac{dQ^*}{dt} \approx -\frac{AM^*}{L}\frac{\delta^3}{2\gamma} \exp\left(-\frac{(2\gamma)^{2/3}}{\delta}\cdot BL\right). 
\end{equation}
Note that $\delta, M^*$ and $\gamma$ are all functions of $t$. We immediately observe that if $\delta$ is small, which corresponds to near-neutral limit, the exponential factor is near unity, but the charge loss rate remains small due to the $\delta^3$ factor. 

In general, it suffices to show that 
\begin{equation}
\frac{\delta^3}{2\gamma} \exp\left(-\frac{(2\gamma)^{2/3}}{\delta}\cdot BL\right) \leq \exp\left(-BL\right).
\end{equation}
Although $\delta$ and $\gamma$ are not independent [$\delta$ increases as $\gamma$ decreases], it is clear that with $\delta \in [0,1]$ and $\gamma \in [1/2,1]$, we must have the upper bound $\delta^3/\gamma \leq 2$. Thus
\begin{equation}
\frac{\delta^3}{2\gamma}\left(e^{-BL}\right)^{\frac{(2\gamma)^{2/3}}{\delta}} \leq \left(e^{-BL}\right)^{\frac{(2\gamma)^{2/3}}{\delta}} \leq e^{-BL}.
\end{equation}
The last inequality follows from 
\begin{equation}
1 \leq \frac{(2\gamma)^{\frac{2}{3}}}{\delta} < \infty
\end{equation}
and the fact that $0 < \exp(-BL) \leq 1$. 

Thus, starting with the same initial mass, the initial charge loss rate for a non-extremal black hole is indeed smaller than that of the extremal black hole. Since mass is monotonically decreasing, the rate for charge loss remains low throughout the evolution.

Therefore, we have the following result:

\begin{proposition}
For any initial mass in the regime of validity of the model\footnote{It must be emphasized that, for asymptotically flat Reissner-Nordstr\"om black holes, HW's analysis \emph{eventually} breaks down when mass drops below certain level. In the case of charged flat black holes however, the evolution \emph{always} stays within the regime of validity, since the model requires large \emph{fixed} $L$, not large $M$.}, and independent of both the compactification parameter $K$ and the [normalized] charge-to-mass ratio $Q^*/(wM^*)$, the charge loss rate of a charged flat black hole, which is given by the metric $g$(AdSRN[k=0]) in Eq.(\ref{ARRGH}), is [practically] zero.
\end{proposition}

Thus, the reason why electrical charge stays almost constant is due to the fact that we are dealing with large AdS length scale $L$, and the fact that $L$ appears in such a way in the Schwinger formula as to conspire to suppress charge production by an \emph{enormously large} exponential factor. Note that this behavior is \emph{not} present in asymptotically flat case, which, as we have seen in the previous section, discharges in a ``reasonably short'' time scale.

\section* {\large{\textsf{4. Conclusion: Charge Loss Inefficiency Leads to Extremal Attractor}}}

We have explained why AdS-Reissner-Nordstr\"om black holes with flat horizon [of either planar or toral topology] and large $L$ \emph{practically} have constant charge, and thus as mass continues to evaporate away, the black holes inevitably evolve towards extremal limit, i.e, \emph{the extremal limit is an attractor}. We also explained why such behavior, which is completely different from asymptotically flat charged black holes, is not inconsistent with the latter. Indeed, while setting $L \to \infty$ in a geometry that corresponds to AdS-Reissner-Nordstr\"om black hole with horizon having spherical topology does recover the same qualitative behavior found in asymptotically flat case, setting $L \to \infty$ in charged flat black hole spacetimes does \emph{not}. The latter simply has different topology, and we cannot pass from one topology to another by taking limit. 

Due to the charge loss rate $dQ^*/dt$ remaining small throughout the evaporation of charged flat black holes in the large $L$ regime, they are all driven towards extremality as they steadily lose mass. Eventually, when these black holes reach around 92\% of the extremal charge, brane-pair production instability \cite{kn:seiberg, KPR} is triggered and they are destroyed, as was argued in \cite{omc}.

\addtocounter{section}{1}
\section*{\large{\textsf{Acknowledgement}}}
Yen Chin Ong and Pisin Chen thank Keisuke Izumi and Shou-Huang Dai for fruitful discussions. Yen Chin Ong also thanks Brett McInnes for useful advice. 
This work is supported by Taiwan's National Center for Theoretical Sciences [NCTS], Taiwan National Science Council [NSC], and the Leung Center for Cosmology and Particle Astrophysics [LeCosPA] of National Taiwan University.

\end{document}